\documentclass[11pt,twoside]{article}
\usepackage{./asp2014}

\aspSuppressVolSlug
\resetcounters

\def\teff{\mbox{$T_{\rm eff}$}}
\def\logg{\mbox{log {\it g}}}

\begin{document}

\title{Helium at white dwarf photospheric conditions: preliminary laboratory results}
\author{M. Schaeuble,$^1$ R. E. Falcon,$^2$ T. A. Gomez,$^{1,2}$ D. E. Winget,$^1$ M. H. Montgomery,$^1$ and J. E. Bailey$^2$}
\affil{$^1$Department of Astronomy and McDonald Observatory, The University of Texas at Austin, Austin, TX 78712, USA; 
\email{mschaeu, gomezt, dew, mikemon@astro.as.utexas.edu}}
\affil{$^2$Sandia National Laboratories, Albuquerque, NM 87185, USA; \email{refalco, jebaile@sandia.gov}}

\paperauthor{M. Schaeuble}{mschaeu@astro.as.utexas.edu}{}{University of Texas at Austin}
{Department of Astronomy and McDonald Observatory}{Austin}{Texas}{78712}{USA}

\paperauthor{R. E. Falcon}{refalco@sandia.gov}{}{Sandia National Laboratories}{}{Albuquerque}{New Mexico}{87185}{USA}

\paperauthor{T. A. Gomez}{gomezt@astro.as.utexas.edu}{}{University of Texas at Austin}
{Department of Astronomy and McDonald Observatory}{Austin}{Texas}{78712}{USA}

\paperauthor{D. E. Winget}{dew@astro.as.utexas.edu}{}{University of Texas at Austin}
{Department of Astronomy and McDonald Observatory}{Austin}{Texas}{78712}{USA}

\paperauthor{M. H. Montgomery}{mikemon@astro.as.utexas.edu}{}{University of Texas at Austin}
{Department of Astronomy and McDonald Observatory}{Austin}{Texas}{78712}{USA}

\paperauthor{J. E. Bailey}{jebaile@sandia.gov}{}{Sandia National Laboratories}{}{Albuquerque}{New Mexico}{87185}{USA}

\begin{abstract}
We present preliminary results of an experimental study exploring helium at photospheric conditions of white dwarf stars. 
These data were collected at Sandia National 
Laboratories' Z-machine, the largest x-ray source on earth. Our helium results could have many 
applications ranging from validating current DB white dwarf model atmospheres to providing accurate
He pressure shifts at varying temperatures and densities. In a much broader context, these helium data 
can be used to guide theoretical developments in new continuum-lowering models for two-electron atoms. 
We also discuss future applications of our updated experimental design, which enables us to sample a 
greater range of densities, temperatures, and gas compositions. 
\end{abstract}

\section{Introduction}
White dwarfs (WDs) represent the endpoint of stellar evolution for roughly 97\% of all stars
\citep{kepler_new_2016}. These stellar remnants can be further subcategorized into DAs, 
which have hydrogen-dominated atmospheres, DBs, whose atmospheres are dominated
by helium, and more exotic species such as DQs, which exhibit atmospheres
almost completely dominated by carbon \citep{sion_proposed_1983}.

Due to their relative physical simplicity, compact size, high mass, and observability in 
most parts of the electromagnetic spectrum, 
WD stars have served as laboratories 
for many different areas of astronomy and physics. They have 
been used as galactic and cosmic chronometers \citep[e.g.,][]{winget_independent_1987,gianninas_ultracool_2015}, 
as a tool for deriving initial-final stellar mass relations \citep[e.g.,][]{cummings_initial-final_2015, cummings_two_2016}, 
as a site to study the compositions of extrasolar planets \citep[e.g.,][]{zuckerman_chemical_2007, farihi_alma_2014}, 
and as a gravitational-field source to constrain our understanding of general relativity \citep{onofrio_search_2014}. 
Additionally, we can use WDs to examine aspects of electroweak theory \citep{winget_strong_2004}, 
as well as our understanding of crystallization in dense Coulomb plasmas
\citep{winget_physics_2009}.

For the purposes outlined above, accurate effective temperature (\teff) and 
surface gravity (\logg) values of these stars
are needed. Using the so-called spectroscopic method, which involves fitting a model atmosphere to an
observed stellar spectrum (see \citealt{bergeron_spectroscopic_1992}, \citealt{tremblay_spectroscopic_2009},
and \citealt{koester_white_2010} for more details), observers have derived these parameters for over 30,000 WDs.
While this method has been proven to be very precise (e.g., $\delta\teff / \teff \sim 5 \% $, $\delta\logg / \logg \sim1 \% $), 
its accuracy is questionable. 
Several studies \citep{falcon_gravitational_2010, falcon_gravitational_2012, barstow_hubble_2005}
have shown that \logg \ values derived using this spectroscopic method do not agree with 
those derived from other, independent methods.
\cite{falcon_experimental_2013} developed an experimental
platform at Sandia National Laboratories' Z-machine to address these \logg \ discrepancies
by verifying H-Balmer spectral line profiles at temperatures and 
densities applicable to DA WD photospheres.
Advances in our understanding of hydrogen line-shape formation have resulted from
these experimental efforts \citep[e.g.,][]{gomez_effect_2016}.
Here, we present preliminary experimental results, which attempt to address similar problems in DB WDs
\citep{falcon_gravitational_2012, koester_db_2015}.

\section{Experimental setup/results}\label{results}
The fundamentals of our experimental platform are described in \cite{falcon_experimental_2013}, and further
details about Sandia's Z-machine as well as our experimental collaboration are given in \cite{rochau_zapp:_2014}. 
In Fig. \ref{fig_1}, we show data collected during an experiment containing a H/He gas mixture. 
The addition of hydrogen allowed us to use the resulting H$\beta$ line to obtain 
plasma temperatures and densities \citep{falcon_laboratory_2015}, which, for the
shown experiment, are consistent with conditions found in DB WD photospheres.

\articlefigure[width=.8\textwidth]{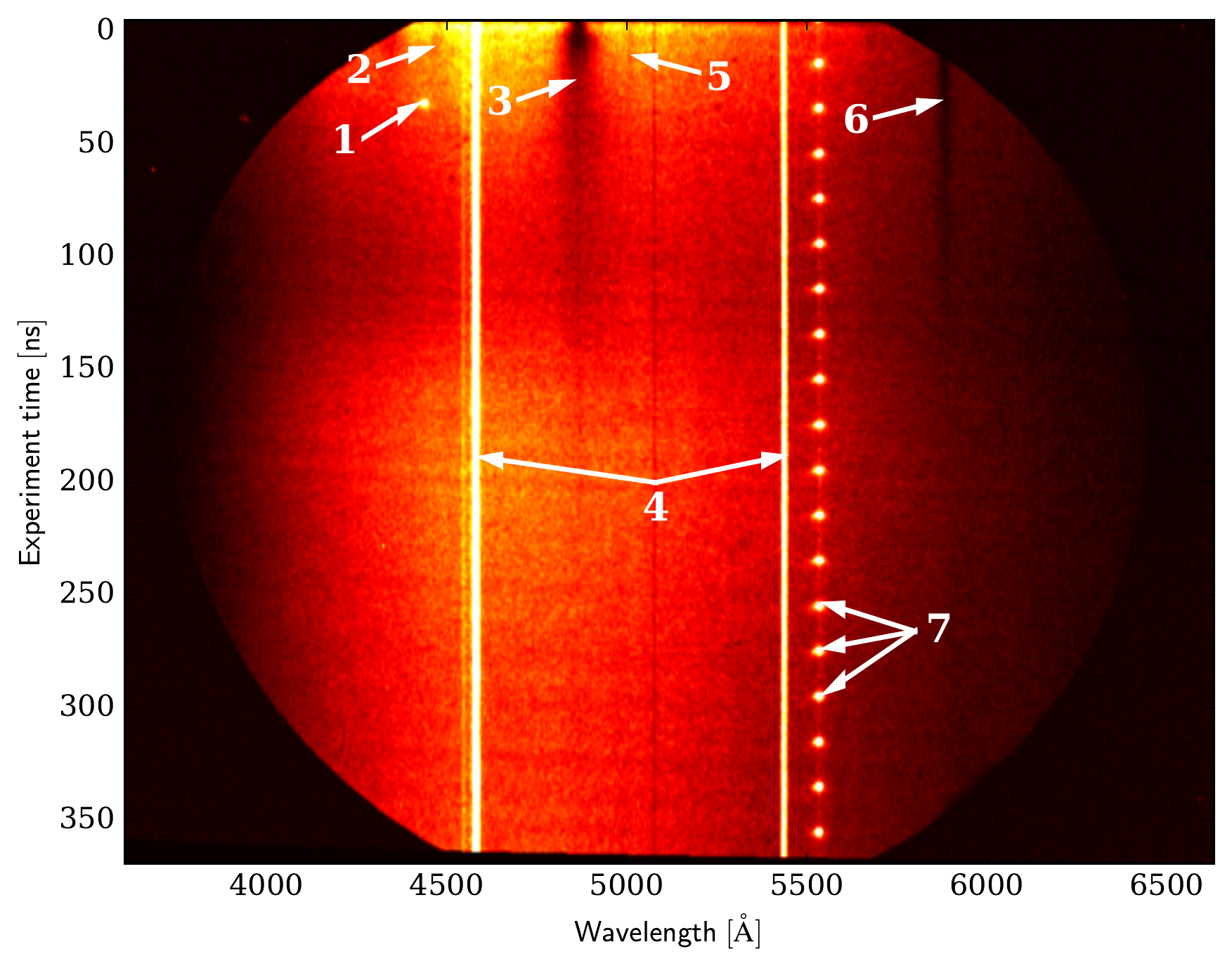}{fig_1}{Time-resolved spectrum of an
experiment containing a mixture of hydrogen and helium. Major features
are identified in the text.}

In Fig. \ref{fig_1}, we identify several prominent emission 
(bright yellow) and absorption (dark red/black) features. The timing impulse, 
labeled `1' in that figure, enables us to determine the beginning of our experiment, 
while the two very bright laser lines at 4579 \AA\ and 5435 \AA\ (labeled as `4') serve as wavelength references. 
The timing comb, identified as
`7', is used to time-resolve our experimental data. The absorption features of our
experimental gas are labeled `2' for the H$\gamma$ line, `3' for the H$\beta$ line, `5' for the \ion{He}{I} feature
at 5015 \AA\ , and `6' for the 5875 \AA\ \ion{He}{I} line. 
The extracted spectra for three different time steps are shown in Fig. \ref{fig_2}, 
where we can clearly see the presence of all the features indicated in Fig. \ref{fig_1}.

\articlefigure[width=.8\textwidth]{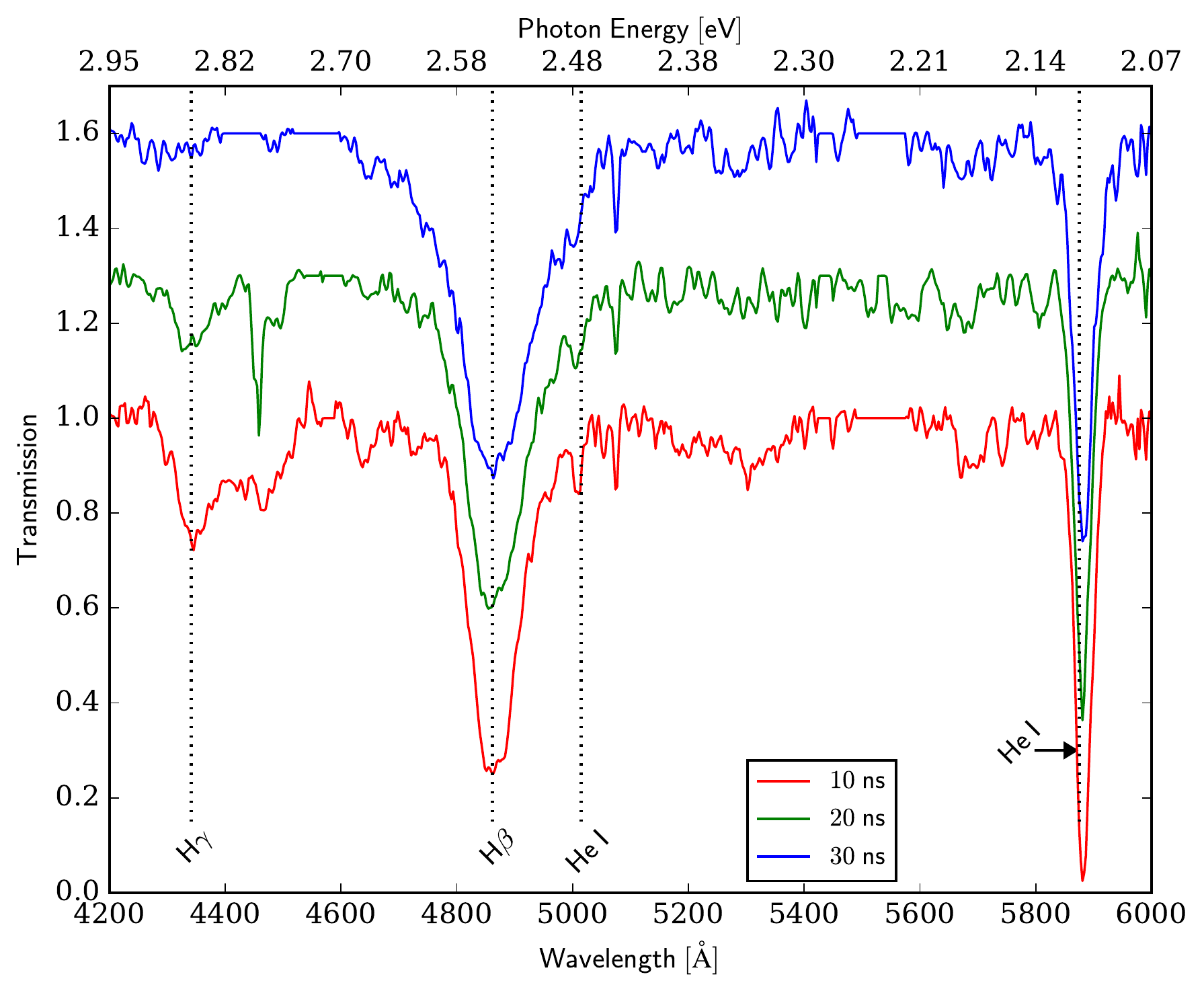}{fig_2}{Spectra extracted at varying time steps. 
The two laser lines (`3' in Fig. \ref{fig_1}) and the timing impulse/comb
(`1' and `6' in Fig. \ref{fig_1}) have been removed for clarity. Major hydrogen and helium spectral
features are identified in the plot. The shown spectra have been offset for clarity.}

\section{Conclusions/Future work}
We have presented preliminary experimental data aimed at addressing persistent
difficulties in model atmospheres for DB WDs. Our results, which are presented in Sec. \ref{results}, show that we
are able to reach DB WD photospheric conditions and thereby explore the problems laid out in
\cite{falcon_gravitational_2012} and  \cite{koester_db_2015}.
The advent of the discovery of hot DQ WDs \citep{dufour_hot_2008} and uncertainties
in their model atmospheres \citep{dufour_stark_2011} have already guided the 
way for future experiments. In fact, we recently performed experiments 
with a significantly altered experimental platform, and a preliminary analysis of the resulting data
shows that we are able to capture relevant atomic species and spectral lines.
Furthermore, an experiment that can validate pure oxygen model atmospheres \citep{kepler_white_2016}
will be developed soon.

\acknowledgements This work was performed at Sandia National Laboratories and is supported by the Laboratory Directed Research 
and Development program. We thank the Z-facility teams and in particular, D. Scogiletti, D. Begay, D. Bliss, 
K. Shelton, A. York, and P. Lake for experimental support. Sandia is a multiprogram laboratory operated by 
the Sandia Corporation, a Lockheed Martin Company, for the United States Department of Energy under 
contract DE-AC04-94AL85000. We also thank A. Wooton for for championing our fundamental science 
research efforts. M.S., M.H.M, and D.E.W. acknowledge support from the United States Department of 
Energy grant under DE-SC0010623. T.A.G. acknowledges support from the National Science 
Foundation Graduate Research Fellowship under grant DGE-1110007.  

\bibliographystyle{asp2014}
\bibliography{EuroWD_2016}  

\begin{thebibliography}{}
\expandafter\ifx\csname natexlab\endcsname\relax\def\natexlab#1{#1}\fi

\bibitem[{Barstow {et~al.}(2005)Barstow, Bond, Holberg, Burleigh, Hubeny, \&
  Koester}]{barstow_hubble_2005}
Barstow, M.~A., Bond, H.~E., Holberg, J.~B., {et~al.} 2005, MNRAS, 362, 1134

\bibitem[{Bergeron {et~al.}(1992)Bergeron, Saffer, \&
  Liebert}]{bergeron_spectroscopic_1992}
Bergeron, P., Saffer, R.~A., \& Liebert, J. 1992, ApJ, 394, 228

\bibitem[{Cummings {et~al.}(2015)Cummings, Kalirai, Tremblay, \&
  Ramirez-Ruiz}]{cummings_initial-final_2015}
Cummings, J.~D., Kalirai, J.~S., Tremblay, P.-E., \& Ramirez-Ruiz, E. 2015,
  ApJ, 807, 90

\bibitem[{Cummings {et~al.}(2016)Cummings, Kalirai, Tremblay, \&
  Ramirez-Ruiz}]{cummings_two_2016}
---. 2016, ApJ, 818, 84

\bibitem[{Dufour {et~al.}(2011)Dufour, Ben~Nessib, Sahal-Br{\' e}chot, \&
  Dimitrijevi{\' c}}]{dufour_stark_2011}
Dufour, P., Ben~Nessib, N., Sahal-Br{\' e}chot, S., \& Dimitrijevi{\' c}, M.~S.
  2011, Bal. Ast., 20, 511

\bibitem[{Dufour {et~al.}(2008)Dufour, Fontaine, Liebert, Schmidt, \&
  Behara}]{dufour_hot_2008}
Dufour, P., Fontaine, G., Liebert, J., Schmidt, G.~D., \& Behara, N. 2008, ApJ,
  683, 978

\bibitem[{Falcon {et~al.}(2015)Falcon, Rochau, Bailey, Gomez, Montgomery,
  Winget, \& Nagayama}]{falcon_laboratory_2015}
Falcon, R.~E., Rochau, G.~A., Bailey, J.~E., {et~al.} 2015, ApJ, 806, 214

\bibitem[{Falcon {et~al.}(2010)Falcon, Winget, Montgomery, \&
  Williams}]{falcon_gravitational_2010}
Falcon, R.~E., Winget, D.~E., Montgomery, M.~H., \& Williams, K.~A. 2010, ApJ,
  712, 585

\bibitem[{Falcon {et~al.}(2012)Falcon, Winget, Montgomery, \&
  Williams}]{falcon_gravitational_2012}
---. 2012, ApJ, 757, 116

\bibitem[{Falcon {et~al.}(2013)Falcon, Rochau, Bailey, Ellis, Carlson, Gomez,
  Montgomery, Winget, Chen, Gomez, \& Nash}]{falcon_experimental_2013}
Falcon, R.~E., Rochau, G.~A., Bailey, J.~E., {et~al.} 2013, HEDP, 9, 82

\bibitem[{Farihi {et~al.}(2014)Farihi, Wyatt, Greaves, Bonsor, Sibthorpe, \&
  Panić}]{farihi_alma_2014}
Farihi, J., Wyatt, M.~C., Greaves, J.~S., {et~al.} 2014, MNRAS, 444, 1821

\bibitem[{Gianninas {et~al.}(2015)Gianninas, Curd, Thorstensen, Kilic,
  Bergeron, Andrews, Canton, \& Agüeros}]{gianninas_ultracool_2015}
Gianninas, A., Curd, B., Thorstensen, J.~R., {et~al.} 2015, MNRAS, 449, 3966

\bibitem[{Gomez {et~al.}(2016)Gomez, Nagayama, Kilcrease, Montgomery, \&
  Winget}]{gomez_effect_2016}
Gomez, T.~A., Nagayama, T., Kilcrease, D.~P., Montgomery, M.~H., \& Winget,
  D.~E. 2016, Phys. Rev. A, 94, 022501

\bibitem[{Kepler {et~al.}(2016{\natexlab{a}})Kepler, Koester, \&
  Ourique}]{kepler_white_2016}
Kepler, S.~O., Koester, D., \& Ourique, G. 2016{\natexlab{a}}, Science, 352, 67

\bibitem[{Kepler {et~al.}(2016{\natexlab{b}})Kepler, Pelisoli, Koester,
  Ourique, Romero, Reindl, Kleinman, Eisenstein, Valois, \&
  Amaral}]{kepler_new_2016}
Kepler, S.~O., Pelisoli, I., Koester, D., {et~al.} 2016{\natexlab{b}}, MNRAS,
  455, 3413

\bibitem[{Koester(2010)}]{koester_white_2010}
Koester, D. 2010, MmSAI, 81, 921

\bibitem[{Koester \& Kepler(2015)}]{koester_db_2015}
Koester, D., \& Kepler, S.~O. 2015, A\&A, 583, A86

\bibitem[{Onofrio \& Wegner(2014)}]{onofrio_search_2014}
Onofrio, R., \& Wegner, G.~A. 2014, ApJ, 791, 125

\bibitem[{Rochau {et~al.}(2014)Rochau, Bailey, Falcon, Loisel, Nagayama,
  Mancini, Hall, Winget, Montgomery, \& Liedahl}]{rochau_zapp:_2014}
Rochau, G.~A., Bailey, J.~E., Falcon, R.~E., {et~al.} 2014, Phy. of Plasmas,
  21, 056308

\bibitem[{Sion {et~al.}(1983)Sion, Greenstein, Landstreet, Liebert, Shipman, \&
  Wegner}]{sion_proposed_1983}
Sion, E.~M., Greenstein, J.~L., Landstreet, J.~D., {et~al.} 1983, ApJ, 269, 253

\bibitem[{Tremblay \& Bergeron(2009)}]{tremblay_spectroscopic_2009}
Tremblay, P.-E., \& Bergeron, P. 2009, ApJ, 696, 1755

\bibitem[{Winget {et~al.}(1987)Winget, Hansen, Liebert, van Horn, Fontaine,
  Nather, Kepler, \& Lamb}]{winget_independent_1987}
Winget, D.~E., Hansen, C.~J., Liebert, J., {et~al.} 1987, ApJL, 315, L77

\bibitem[{Winget {et~al.}(2009)Winget, Kepler, Campos, Montgomery, Girardi,
  Bergeron, \& Williams}]{winget_physics_2009}
Winget, D.~E., Kepler, S.~O., Campos, F., {et~al.} 2009, ApJL, 693, L6

\bibitem[{Winget {et~al.}(2004)Winget, Sullivan, Metcalfe, Kawaler, \&
  Montgomery}]{winget_strong_2004}
Winget, D.~E., Sullivan, D.~J., Metcalfe, T.~S., Kawaler, S.~D., \& Montgomery,
  M.~H. 2004, ApJ, 602, L109

\bibitem[{Zuckerman {et~al.}(2007)Zuckerman, Koester, Melis, Hansen, \&
  Jura}]{zuckerman_chemical_2007}
Zuckerman, B., Koester, D., Melis, C., Hansen, B.~M., \& Jura, M. 2007, ApJ,
  671, 872

\end{thebibliography}

\end{document}